\documentclass[11pt,a4paper]{article}
\usepackage[margin=2.5cm]{geometry}
\usepackage[utf8]{inputenc}
\usepackage[T1]{fontenc}
\usepackage{lmodern}
\usepackage{amsmath,amssymb}
\usepackage{graphicx}
\usepackage{authblk}
\usepackage[numbers,sort&compress]{natbib}
\usepackage[hidelinks]{hyperref}
\usepackage{microtype}
\title{A Simulation Testbed for Self-Governing Human-Machine Institutions}
\author[1]{Callum Berger}
\author[2]{Jeremy Pitt}
\author[1,*]{Stefan Sarkadi}
\affil[1]{APEx Defence and Security Institute, University of Lincoln, Lincoln, UK}
\affil[2]{Department of Electrical and Electronic Engineering, Imperial College London, London, UK}
\affil[*]{Corresponding author: ssarkadi@lincoln.ac.uk}
\date{}

\begin{document}
\maketitle

\begin{abstract}
AI agents no longer act alone: they cooperate, coordinate and compete within and between groups, increasingly alongside humans who are in, on or out of ``the loop''. Whether such mixed groups can govern themselves is an open question about machine behaviour and human-computer interaction that needs a testbed for hybrid institutions. We present Megabike, a multi-agent simulator and serious game with interdependent arenas for common-pool resource management, cooperative survival and constitutional choice. Riders sharing a megabike must agree their arrangements for steering, pedalling and distributing resources, and their political regime, while negotiating with rival bikes and avoiding an existential threat. Every rule is enforced by the simulator, every action is logged, and any seat can be taken by a human or a machine, so governance can be studied under well-defined metrics. We demonstrate Megabike with forty sessions in which language-model riders with cooperative, lazy, subversive or opposing dispositions chose regimes, elected leaders and expelled one another. The results show that different combinations of agent archetypes could produce unexpected and counterproductive outcomes. We conclude with extensions for systematic experiments mixing human and computational agents.
\end{abstract}

\noindent\textbf{Keywords:} self-governance, societies, LLMs, human-machine institutions, machine behaviour

\section{Introduction}

AI agents are no longer deployed one at a time. Systems built around large
language models (LLMs) are increasingly organised as teams of agents that
split tasks between them, negotiate over shared resources, and take decisions
on behalf of the people who deploy them. The question to ask of them is no
longer only whether an agent can complete a task, but what happens when a group
of such agents has to decide, by itself, how it is going to be run: can groups
of AI agents govern themselves, and if so, how well?

Multi-agent systems (MAS) research has studied this question under the label of
self-organising or self-governing systems \cite{ostrom1990,pitt2012axiom,pitt2021book}.
Ostrom showed that communities can manage shared resources over long periods
without privatisation or external control if their institutions follow a small
number of design principles, among them clear boundaries, participation of those
affected in making the rules, monitoring, and graduated sanctions
\cite{ostrom1990}. She also distinguished three levels of rules, namely operational
rules about using the resource, collective-choice rules about changing the
operational rules, and constitutional rules about who belongs to the group and
how the collective-choice rules themselves can change. Pitt and colleagues
turned these principles into executable specifications for MAS and tested them
in simulation \cite{pitt2012axiom,pitt2014distributive,pitt2021book}, building
on formalisations of norm-governed societies and institutionalised power
\cite{jonessergot1996,artikis2009}. Constitutional choice is the hardest level because it concerns who rules and how rulers are removed, and its effects appear only over long time scales \cite{mertzani2024cohesion}. Ober's thought experiment \emph{Demopolis}
splits the founding of an institution into self-selection of members,
entrenchment of the initial rules, and operationalisation \cite{ober2017}, and
Pitt and Ober's democracy-by-design translates this into design principles for
MAS, including that democratic processes should not be short-cut
\cite{pittober2018}. In all of this work, however, the agents were designed by
the experimenters. The agents' preferences were utility functions, their options were
enumerated in advance, and their ``deliberation'' was a pre-established protocol rather than a set of rules reached through debate and deliberation.

LLM-based agents can easily act as if they debate. They have been shown to simulate small communities
\cite{park2023generative}, to negotiate at human level when coupled with a
strategic reasoner \cite{cicero2022}, and to manage, or rather fail to manage,
a common-pool resource \cite{piatti2024govsim}. But in this line of work the
rules of the game are fixed by the experimenter, and what is measured is task
performance, cooperation rates, or the number of ethical violations
\cite{leibo2021meltingpot,pan2023machiavelli}. What is not measured is whether
the agents can found an institution, run it under pressure, notice when it
stops working, and change it. 

This is rather important, especially since our societies are becoming hybrid, in which humans and machines
increasingly adapt to each other's behaviour in a shared Infosphere
\cite{floridi2014,sarkadi2024selfgov}, and deception is one of the behaviours
that spreads most easily in such societies
\cite{sarkadi2021evolution,sarkadi2023arms}. LLMs produce fluent text that is
indifferent to truth \cite{hicks2024bullshit}, their users under- or
over-recognise this as deception \cite{zhan2025banal}, and their cheapness
undermines the signals, such as effort, that humans rely on when deciding whom
to trust \cite{lewismarsh2022,powers2025trustllm}. A group that governs itself
through deliberation is therefore also a group whose governance can be attacked
through deliberation \cite{piazza2025cpr}. Whether such groups can nonetheless self-govern cannot be settled a priori; it has to be observed, and analysed a posteriori in the spirit of Machine Behaviour \cite{rahwan2019machine}. On their own, LLMs are unreliable planners that cannot be trusted to respect constraints they have agreed to, though they work well as generators and critics inside a loop that verifies their output \cite{kambhampati2024llmmodulo}; hence neurosymbolic architectures, which pair neural flexibility with inspectable, enforceable symbolic structures \cite{garcezlamb2023}, a separation long familiar in MAS frameworks such as JaCaMo, which keep agent, environment and organisation apart so that institutional rules bind independently of any agent \cite{boissier2013jacamo}.

To test self-governance we need an environment in which (i) decisions about
rules have consequences the agents cannot escape, (ii) agents decide at every
level of governance, from who invests effort to who rules, (iii) cooperative and
adversarial incentives are present at the same time, (iv) the agreed rules are
explicit and inspectable, and (v) different kinds of agents, including
eventually humans, can play the same roles. The Megabike scenario has all of these
properties. It comes out of a series of cooperative survival games in which no
agent survives unless the group survives
\cite{scott2022motives,scottpitt2023,pitt2025tmt}. Scott \emph{et al.} introduced it to study whether costly deliberation could be replaced by social contracts agreed once, and found that the mutability of rules mattered more than the abundance of resources: bikes with immutable rules survived roughly half as long \cite{scott2024megabike}. Mertzani \emph{et al.} formalised a regime change game between rule by one, by few, by many or by an external authority, and used Megabike to study how explaining constitutional choice relates to cohesion \cite{mertzani2024cohesion,mertzani2023theta}. Two things distinguish Megabike from the usual gridworld testbeds. The first is that institutional and physical actions depend on each other, since a leader's declared target changes what counts as compliant pedalling and only the steerer can turn the bike \cite{artikis2009}. The second is that the rules are an object in the environment, evaluated by the simulator, so non-compliance can be detected and sanctioned without coercion. Previous Megabike studies used hand-coded riders and point to human-in-the-loop experiments as the next step.

In this article we present Megabike as a serious-game testbed for the
self-governance of hybrid institutions \cite{ober2017}, and report a first set of experiments in which the riders are neurosymbolic LLM agents whose world model, voting rules and psychological profile are symbolic, and whose arguments are generated in natural language. The experiments involve only artificial agents, but Megabike was designed to be played by humans and machines together \cite{scott2024megabike,mertzani2024cohesion}, and the all-machine case is the first step towards studying hybrid societies and
human-machine teams on the same platform \cite{patel2025hmt}.

\section{The Megabike Platform}
\label{sec:platform}

\subsection{The scenario}

\begin{figure}[t]
\centering
\includegraphics[width=0.75\textwidth]{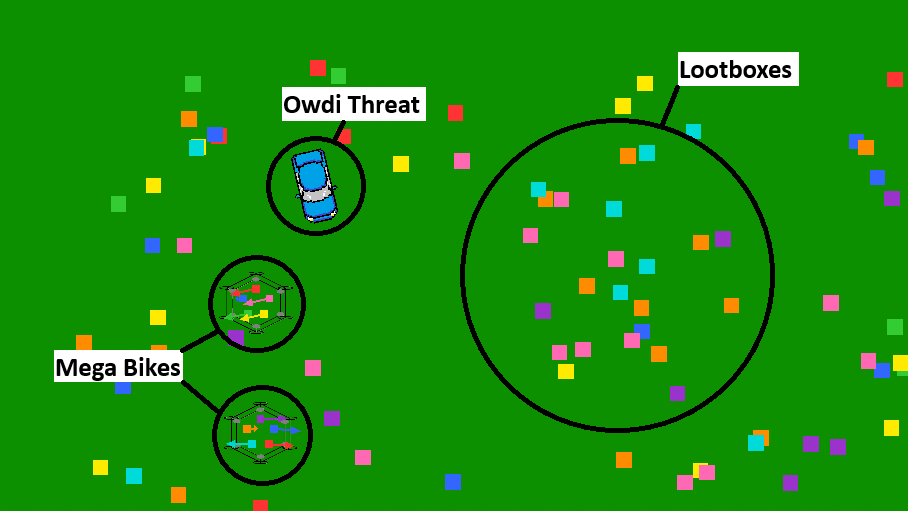}
\caption{The Megabike scenario. Five riders share one frame, each with their own
pedals, brakes and steering wheel, so the bike moves according to their combined
actions. Institutional roles sit on top of physical capability: a leader
declares the target lootbox and the required effort, and only the steerer turns
the wheel. Lootboxes supply the energy on which survival depends and are
contested between bikes; the existential threat destroys any bike it reaches.}
\label{fig:scenario}
\end{figure}

A megabike is a bicycle for several riders, inspired by real multi-rider
bikes, but where every seat has its own pedals, brakes and steering
wheel. The bike moves according to the combined effect of what all the riders
do. If some pedal while others brake, energy is wasted and the bike goes
nowhere; if everyone steers at once, the bike swerves. A group of autonomous
agents on a megabike therefore shares a single body that none of them controls
alone, and has to agree on social arrangements that settle who decides
direction, effort and target, how rewards are shared, and who takes which role
\cite{scott2024megabike}. Figure~\ref{fig:scenario} illustrates the scenario.

Megabikes move around a gridworld that contains lootboxes and an existential
threat (in the current build, a hostile vehicle called the Owdi). Lootboxes hold the energy that agents need to survive; they appear at
random locations, remain for a limited time, and disappear when collected by
any bike. The threat moves through the world and destroys any megabike it
reaches. Several megabikes can be active at once, so lootboxes are shared
within a bike but contested between bikes. Each agent wants to maximise its
own utility, in terms of energy gained, time survived, and the esteem of the
other riders for its contribution. However, the energy spent pedalling by a single agent does not
increase its likelihood of survival.

\subsection{Energy and physical actions}

Each agent has a limited amount of energy, which decreases every round and is
replenished only from lootboxes. Every round, each agent decides how hard to
pedal or brake and how to steer, at an energy cost proportional to the action.
The environment sums pedalling and braking into the bike's speed and turns the
bike according to whichever steering input the rules treat as authoritative:
the combined steering of all riders, or that of the agent holding the steerer
role. A collected lootbox goes into the bike's common stock and is divided
according to the allocation rule in force. An agent whose energy runs out is
eliminated and its seat becomes vacant; a bike that loses all its riders, or is
caught by the threat, is destroyed. Settings for the experiment reported here are given in Section~\ref{sec:experiments} and in Appendix~S2.

\subsection{Social arrangements}

The social arrangements of a megabike are a set of rules spanning Ostrom's
three levels \cite{ostrom1990}. Following
\cite{scott2024megabike,mertzani2024cohesion}, we group them into the decision
arenas in Table~\ref{tab:arenas}. In each arena the agents agree on a rule type and its parameters. Every constitution is an explicit
object bound to the bike, so it can be recorded, inspected and compared across
bikes and over time.

The arenas depend on each other. An allocation rule that rewards effort
presupposes a monitoring rule that can measure it, and a leader who can
declare the target is only useful if the rules make that declaration binding.
The arrangements also include rules about their own mutability, i.e., the
agents can declare some rules immutable and must then agree on how a rule is
entrenched or released again \cite{suber1990}. This is what allows us to
observe democratic backsliding. A bike that hands power to a single leader
under threat and has no procedure for taking it back has not made an error
that the environment punishes immediately, since centralisation is often
efficient in the short run, but it has lost a capacity it will need when
conditions change \cite{graeberwengrow2021,mertzani2024cohesion}. At the top
of the arrangements is a political regime: rule by one, by few, by many, or
by an external authority \cite{ober2017}. In the implementation reported here, only rule by one and rule by many are available. Moving between regimes is the regime
change game, and each move promotes some values and demotes others (represented by the colours of the lootboxes)
\cite{mertzani2024cohesion}.

\begin{table}[t]
\centering
\caption{Decision arenas of the Megabike constitutional framework. In each
arena the riders agree on a rule type and its parameters; the resulting
constitution is bound to the bike and visible to every rider. Level follows
Ostrom's hierarchy of rules \cite{ostrom1990}, with a socio-political level for
role and decision rules and a meta-constitutional level for crisis rules.}
\label{tab:arenas}
\small
\begin{tabular}{@{}p{0.16\textwidth}p{0.27\textwidth}p{0.33\textwidth}p{0.11\textwidth}@{}}
\hline
\textbf{Arena} & \textbf{Question settled} & \textbf{Examples of rule types and parameters} & \textbf{Level} \\
\hline
Regime & Who rules? & one / few / many / external authority; how the ``few''
are chosen & constitutional \\
Mutability & Which rules may change, and how? & mutable vs.\ immutable rules;
how a rule is entrenched or released; majority thresholds & constitutional \\
Membership & Who is bound by the rules? & admission by invitation or vote;
exclusion with pre-conditions and appeal & constitutional \\
Role assignment & Who holds which power? & leader, steerer, auditor; chosen by
vote, criteria, rotation or incumbency; how they are removed & socio-political \\
Decision method & How are collective choices made? & leader decides, plurality,
alternative vote, Borda count, consensus; quorum; time allowed for debate &
socio-political \\
Monitoring \& sanctions & How is compliance checked and enforced? & audit
method and cost; graduated sanctions; sanction for false reporting &
socio-political \\
Target selection & Which lootbox, and with what effort? & constraints on
distance and payoff; required pedalling effort & operational \\
Allocation & How is loot divided? & equal shares, by need, by contribution, by
esteem, at the leader's discretion & operational \\
Crisis response & What changes when the threat is near? & trigger; temporary
centralisation; when and how power reverts & meta-constitutional \\
\hline
\end{tabular}
\end{table}

\subsection{Phases of the game}

A Megabike game runs over a number of iterations, each following the three
stages of Ober's \emph{Demopolis} \cite{ober2017,scott2024megabike}:
\begin{enumerate}
\item \emph{Self-selection.} Agents negotiate which megabike to join. At the
  start of a game this is a decision under a veil of ignorance
  \cite{rawls1971}; as agents observe each other over iterations, it becomes a
  decision based on reputation.
\item \emph{Entrenchment.} The agents on a bike negotiate their social
  arrangements, i.e., a choice for every arena in Table~\ref{tab:arenas}, and
  assign roles. What is agreed here governs the rest of the iteration.
\item \emph{Operationalisation.} The bike plays a sequence of rounds. In each
  round the agents apply their rules to decide on a target lootbox and the
  required effort, each agent chooses its own physical action, the environment
  moves the bikes and the threat, and any collected lootbox is divided
  according to the allocation rule. The round ends with monitoring, sanctions,
  and any admissions to or exclusions from the bike.
\end{enumerate}
An iteration ends after a fixed number of rounds, or when the bike is destroyed by the Owdi, or the agents on it run out of energy.

\subsection{Dilemmas}

A Megabike game composes several dilemmas \cite{scott2024megabike}: a
collective action dilemma, since free-riding is individually optimal but a
bike of free-riders is stationary and destroyed; a resource allocation dilemma,
since every rule for dividing a lootbox favours some claims over others
\cite{rescher1966,pitt2014distributive}; a balloon debate, in which an agent
facing exclusion must argue for staying, which is where persuasive but
unreliable speech pays most; and competition between bikes for finite
lootboxes. Cooperation within the bike and competition between bikes are live
at the same time.

\subsection{Agents and measurements}

The platform does not care what occupies a seat. A hand-coded agent, a learning
agent, an LLM agent or a human all have to choose or accept a bike, take part
in deliberation and voting, act physically every round, and respond to audits.
Nothing forces these to be truthful or mutually consistent, which is what makes
monitoring and sanctions meaningful. Because the rules are explicit and the
environment adjudicates compliance, governance can be measured directly:
survival and welfare (rounds survived, energy accumulated, lootboxes won
against rival bikes); distribution (how unequal allocations are and how they
relate to contribution); constitutional dynamics (the sequence of regimes a
bike passes through, how often rules are amended, whether centralisation
adopted in a crisis persists); and the cost and integrity of deliberation
(deliberation cost, non-compliance with one's own votes, misleading claims). The last connects Megabike to our earlier
work on deception in self-governing hybrid societies
\cite{sarkadi2024selfgov,zhan2025banal,powers2025trustllm}.

\section{Experimental Demonstration}
\label{sec:experiments}

\paragraph{Design and conditions} The experiment is between-sessions with four
conditions and ten sessions per condition. In every session ten LLM agents ride
two bikes, five per bike. Four seats on each bike hold one rider of each
personality: cooperative, lazy, subversive and opposing. The fifth seat holds a
second rider of the personality that defines the condition, so each condition
gives one personality a plurality of two seats out of five on both bikes, with
the other three represented once each, and the two bikes in a session are
structurally matched. The conditions are named after the doubled personality: Cooperative, Lazy, Subversive and Opposing. All other parameters are held constant, and
conditions were run in sequence, cycling through the four and repeating.

\paragraph{Agents} All ten riders are LLM-controlled; there are no human
participants. Each agent runs Qwen3.5-4B-Q4\_K\_M through the LLMUnity library,
which integrates LLMs into the Unity game engine. Riders are identified to one another only as Player 1 to Player 10, and colours are assigned by seat. Each personality is set by a system prompt (cooperative and agreeable;
passive and low-effort; inconsistent and hard to read; contrarian and
argumentative) that governs both the agent's movement controller, i.e., how
closely it follows the bike's agreed rules, and its language. A common set of
communication rules limited turns to two sentences and required a stated
reason. Each agent observed its own identity and colour, every rider's loot
score with the leader marked, who had stated a leaning so far, the last twelve
lines of dialogue, and its permitted answer options. The full prompts,
communication rules and world view are given in Appendix~S1.

\paragraph{Task} A session is one iteration in the sense of
Section~\ref{sec:platform} and lasts up to five rounds of 120 seconds, each round
followed by a deliberation. Riders steer and pedal a shared bike, collecting
coloured lootboxes. In our current build, energy takes the form of stamina, which
pedalling consumes and lootboxes restore, and each rider also accumulates
points, which are visible to every rider and serve as the score. A box gives
one point to each rider aboard and one to the bike, and restores the stamina of
all riders; it is worth more to a rider whose colour matches it. Steering costs
no stamina, but pedalling does. The threat, the Owdi, crosses the arena on fixed
linear paths and, on contact, forces that bike to remove a rider. The remaining
parameters are given in Appendix~S2.

\paragraph{Governance} Before the game begins, each bike votes on whether
decisions will be made by the group or by an elected leader. Under the chosen
regime, the bike agrees two operational rules, which colour of lootbox to chase
and how hard to pedal. A leader holds office for a voted term and controls
steering. At each round end, riders hold three discussion turns in which each
states a leaning on whom to remove from the bike, followed by a binding vote in
which a rider may also vote for nobody; under a leader, the riders plead their
case and the leader decides. Of the arenas in Table~\ref{tab:arenas}, regime, decision method, role assignment, target selection, mutability and membership were put to the riders; allocation was fixed, and no audit rule was instantiated, so riders saw one another's points but not their effort or compliance. The platform allows a removed rider to negotiate a place on the
other bike, offering accumulated points; this transfer option was disabled in
the sessions reported here, so every removal was an elimination.

\paragraph{Hypotheses} Stability on a megabike can be read on its rules, collective effort, governance, individual behaviour and survival. We test two hypotheses.
\textbf{H1:} A megabike of mixed dispositions is unstable: within five rounds it
will amend its rules, change regime or eliminate riders rather than settle into
stable governance. \textbf{H2:} Giving one disposition a plurality moves
stability in the direction of that disposition: a Cooperative plurality raises
effort and survival, a Lazy plurality lowers them, and Subversive or Opposing
pluralities destabilise governance, with more leader regimes, more rule changes
and more eliminations.

\section{Results}

Forty sessions were run, ten per condition, on one frozen build with the same
local model. They produced 9,255 utterances, 33,229 collection events and
64,510 movement samples; 8,454 utterances are model output, 45 are ellipsis
outputs and 756 are turns in which a rider declined to speak. Condition-level tests are Kruskal--Wallis on four groups
of ten sessions ($df=3$ throughout); personality-level measures are
within-session and computed on session-level means; pairwise comparisons are
Holm-corrected; and contrasts chosen after inspection are checked against a
permutation distribution and reported at their selection-adjusted
significance. Full statistical methods are given in Appendix~S3.

\subsection{Session outcomes}

Eighteen sessions entered round five and twelve completed it; the rest ended
when a bike was emptied (25) or one rider remained (3). Cooperative sessions
lasted longest, 4.8 rounds against 4.0--4.1 (Table~\ref{tab:results}a), with
eight of ten entering round five against two Lazy, three Subversive and five
Opposing. As a rank comparison of round counts this falls short ($H=7.11$,
$p=.068$); as a proportion reaching round five it holds ($\chi^2=8.49$, $df=3$,
$p=.037$), and the Cooperative contrast against the remainder sits at $p=.044$
once its post-hoc selection is accounted for. Lazy sessions absorbed 3.4 Owdi
strikes against 1.9 for Cooperative ($H=8.24$, $p=.041$). Eliminations ($p=.147$), deliberations
($p=.199$) and the choice of leader over group vote ($p=.863$) did not differ.

\begin{table}[t]
\centering
\caption{(a) Session outcomes by condition: means over ten sessions (SD). Owdi
strikes are collisions with the hostile vehicle, forcing a rider's removal;
deliberations excludes the two opening deliberations, leader-led deliberations
counts all. (b) Behaviour and language by personality. Effort and compliance are
session-level means (SD); silence is the share of discussion turns in which the
rider declined to speak (all 756 were Lazy); eliminated is the share of its 100
rider-slots removed. Language columns are percentages of non-silent turns
(words: mean turn length). All language features differ at $p<.001$.}
\label{tab:results}
\footnotesize
\begin{tabular}{@{}lccccc@{}}
\multicolumn{6}{@{}l}{\textbf{(a) Session outcomes by condition}}\\
\hline
\textbf{Condition} & \textbf{Rounds} & \textbf{Eliminations} & \textbf{Owdi strikes} & \textbf{Deliberations} & \textbf{Leader-led deliberations} \\
\hline
Cooperative & 4.8 (0.4) & 7.7 (0.9) & 1.9 (1.1) & 9.1 (1.2) & 1.5 (1.0) \\
Lazy        & 4.1 (0.6) & 8.0 (0.5) & 3.4 (1.0) & 8.4 (1.0) & 1.5 (0.8) \\
Subversive  & 4.0 (0.8) & 7.3 (0.7) & 2.7 (1.1) & 7.6 (1.4) & 1.9 (1.0) \\
Opposing    & 4.1 (1.0) & 7.2 (1.1) & 2.8 (1.3) & 8.4 (2.1) & 1.7 (0.9) \\
\hline
\end{tabular}

\vspace{6pt}

{\setlength{\tabcolsep}{3.5pt}\resizebox{\textwidth}{!}{%
\begin{tabular}{@{}lcccccccccccc@{}}
\multicolumn{12}{@{}l}{\textbf{(b) Individual behaviour and language by personality}}\\
\hline
 & \multicolumn{4}{c}{\textbf{Behaviour}} & \multicolumn{7}{c}{\textbf{Discussion turns (\%)}} \\
\cline{2-5}\cline{6-12}
\textbf{Personality} & \textbf{Effort} & \textbf{Compliance} & \textbf{Silence} & \textbf{Eliminated} & \textbf{Words} & \textbf{Reason} & \textbf{Qualify} & \textbf{Agree} & \textbf{Accuse} & \textbf{Defend} & \textbf{Ask} \\
\hline
Cooperative & 0.544 (0.063) & 100\% (0.0)  & 0\%  & 76\% & 18.4 & 66 & 21 & 35 & 5  & 3  & 0  \\
Lazy        & 0.145 (0.013) & 100\% (0.0)  & 65\% & 83\% & 15.6 & 83 & 5  & 30 & 4  & 4  & 0  \\
Subversive  & 0.446 (0.028) & 49.9\% (5.0) & 0\%  & 69\% & 17.4 & 65 & 16 & 11 & 10 & 7  & 1  \\
Opposing    & 0.516 (0.041) & 1.4\% (0.8)  & 0\%  & 74\% & 18.8 & 55 & 3  & 1  & 15 & 13 & 12 \\
\hline
\end{tabular}}}
\end{table}

\subsection{Collective effort}

Pedalling effort differed sharply by condition ($H=23.94$, $p<.001$;
$\eta^2=0.654$): 0.345 of maximum in the Lazy condition against 0.455 in the
Cooperative, with Subversive (0.421) and Opposing (0.422) between. Doubling the
riders who comply with the rules but hardly pedal cut collective output by
roughly a quarter relative to doubling those who pedal fully. Effort declined over the first three rounds in all four conditions, which differ in where they
start ($H=32.47$, $p<.001$ on round one) but not in how fast they fall
($H=4.38$, $p=.224$). Loot fell faster, from 329--400 events per session in
round one to 48--84 in round five. Since loot is the only source of stamina,
the Lazy condition resulted in less effort, a slower bike, less loot and more
strikes. From round four the figures rest on a shrinking, non-random set of sessions, so later changes reflect survival.

\subsection{Individual behaviour}

Each personality behaved as its controller specified
(Table~\ref{tab:results}b): Cooperative and Lazy riders followed the agreed
rules in every sampled frame, Subversive riders in 49.9\% and Opposing riders
in 1.4\%. Effort differed across all four personalities ($H=118.82$,
$p<.001$), with all six pairwise comparisons significant after Holm
correction. Rule-following and effort came apart: Lazy riders steered as agreed
while pedalling at 0.145 of maximum, whereas Opposing riders pedalled at 0.516,
close to the Cooperative 0.544, while steering as agreed in 1.4\% of samples.
Elimination showed no such differentiation. Each personality occupies exactly
100 rider-slots; the counts were Cooperative 76, Lazy 83, Subversive 69 and
Opposing 74 ($\chi^2=5.46$, $df=3$, $p=.141$; permutation $p=.167$). The null is not equivalence; the defensible claim (Appendix~S3.1) is that Opposing and
Cooperative riders were eliminated at 74\% and 76\% despite differing in
compliance by 98.6 percentage points.

\subsection{Deliberation}

Personalities differed on every discussion feature in Table~\ref{tab:results}b
(all $p<.001$). Opposing riders produced the longest turns while justifying
them least, led on accusation (15\%) and self-defence (13\%), and were almost
alone in asking questions (12\% against 0--1\%), rhetorically. Agreement fell from Cooperative (35\%) and Lazy (30\%)
through Subversive (11\%) to Opposing (1\%). Qualification separates the same
extremes, Cooperative 21\% against Opposing 3\%, but inverts the middle,
Subversive 16\% against Lazy 5\%: hedging tracks willingness to be contradicted
rather than cooperativeness. Each personality also has a lexical register (Appendix~S4.2); the Lazy one is almost entirely reported speech (\textit{since},
\textit{said}, \textit{argued}), so its high justification rate (83\%) is
citation rather than argument. A rider saying \textit{``I'll go with Group
since Player 8 just said trusting one person is risky''} registers as agreeing
and reasoning while contributing neither.

\subsection{How riders argue}
\label{sec:discourse}

Three features of the 415 deliberations stand out (full account in Appendix~S4.1).
First, deliberations are not about the event that triggered them: only 1.7\%
of discussions following an Owdi removal refer to the collision, the vehicle or
any threat, marginally fewer than when no collision occurred (2.6\%). Second, a
rider's loot is argued both ways. Of 1,632 turns mentioning a rider's loot
standing, 250 (15\%) treat it as a negative for the team and 166 (10\%) as a
positive, with 60 doing both at once; since a score reflects what the whole bike collected, it supports either narrative. Agents also accuse one another of acts the game makes
impossible, such as lying (109 turns), theft (80), framing or conspiracy (23)
and sabotage (13), and read the output of the fixed allocation rule as evidence
of collusion. Third, agents attach to targets already named rather than to new
behaviour. Of 125 turns citing a named agent's silence, 73 name an agent who
had already spoken in that discussion, and of 3,029 discussion turns, 32\% name
an agent who has already named the speaker, against 23.4\% expected from
targeting rates alone. Under a leader, the removal decision differed from the
leader's earlier stated choice in 38\% of deliberations (38 of 101).
Rule votes, by contrast, were approved almost without exception: of 611
individual ballots on reopening the agreed rules, 607 were against, so none of
the 205 collective reopening votes passed and every bike played to the end
under the rules it adopted in its first deliberation. Justifications treated
whatever happened in the previous round as validation that the rules work,
including a removal the speaker knew had been settled by a coin flip.

\subsection{Power-seeking and reactivity}

Two behaviours emerged that no personality prompt specified. The first is
willingness to seek office. In leader elections, where self-nomination was
permitted but not encouraged, Opposing riders nominated themselves in 81\% of
turns, Subversive in 64\%, Lazy in 37\% and Cooperative in 17\% ($H=54.86$,
$p<.001$; Figure~\ref{fig:sanctions}c): the riders least willing to follow the
group's rules were the most willing to seek the authority to set them. The
second is weak correspondence between stated and enacted position. A rider's
opening discussion leaning matched its binding vote in 41.4\% of cases and its
final leaning in 43.1\%; because votes converge on a single target, 27.1\% of
this agreement arises by coincidence, so discussion predicts the vote above
chance ($p<.001$), but most of the prediction is already in the first leaning
and the deliberation that follows adds under two percentage points.
Personality differences here are mostly an artefact of abstention (Appendix~S5); restricted to votes naming a rider, only Opposing riders separate, at 36\% against 53--57\%: the riders who argued most forcefully were the only group whose words failed to predict their votes.

\subsection{Governance choices}

Bikes overwhelmingly preferred collective decision-making: of 342 deliberations
putting the method to a vote, 276 (81\%) chose a group vote and 66 (19\%)
elected a leader, with no difference between conditions ($H=0.74$, $p=.863$).
The 66 elections produced 116 binding removal decisions over multi-round terms, and 14 (12\%) spared everyone. Of 210 rules agreed, 148
(70\%) sent riders after their own colour and 143 (68\%) set effort to
\textit{Steady}; agreed effort did not differ across conditions ($p=.091$), so
it does not explain the differences in realised effort, and Opposing bikes
agreed to the most demanding rules while realising less effort than
Cooperative ones. Leader-led rounds were less productive: over rounds one to
three, before survivorship biases the comparison, bikes under a leader
collected 26.0 lootboxes per rider per round against 31.4 under group rule
(Mann--Whitney $p=.008$), with lower compliance (0.60 against 0.64) and lower
effort (0.38 against 0.42).

\subsection{Who sanctions whom}

\begin{figure}[t]
\centering
\includegraphics[width=\textwidth]{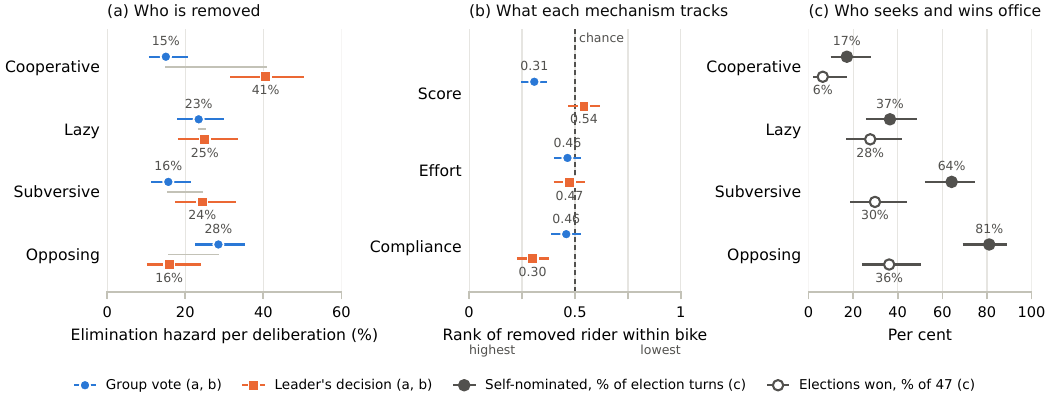}
\caption{Sanctions and office across all forty sessions. (a) Elimination hazard
by personality under group vote (circles) and a leader's decision (squares):
the share of a personality's appearances in a deliberation that reached the
elimination stage which ended in its removal, with 95\% Wilson intervals.
(b) Mean within-bike rank of the removed rider on score, effort and compliance
in the preceding round (0 highest, 1 lowest, dashed line chance), with 95\%
confidence intervals. (c) Share of a personality's election turns in which it
nominated itself (filled) and share of the 47 contested elections it won
(open), with 95\% Wilson intervals.}
\label{fig:sanctions}
\end{figure}

Breaking the 302 eliminations down by mechanism (158 by group vote, 116 by a
leader's decision, 28 from a lone rider being caught) explains the null above,
because the two mechanisms targeted different riders
(Figure~\ref{fig:sanctions}a). Under group vote the hazard per deliberation was
28.5\% for Opposing and 23.4\% for Lazy riders against 15.0\% for Cooperative
and 15.7\% for Subversive ($\chi^2=14.68$, $df=3$, $p=.002$); under a leader it
inverted, to 40.6\% for Cooperative against 25.0\% Lazy, 24.3\% Subversive and
15.9\% Opposing ($\chi^2=17.01$, $df=3$, $p<.001$). Pooled, the two gradients
cancel into the undifferentiated 69--83\% reported above. The reversal follows
from who won office (Figure~\ref{fig:sanctions}c): Opposing riders won 17 of
the 47 contested elections, Subversive 14, Lazy 13 and Cooperative 3 (Opposing
against Cooperative, binomial $p=.003$), and Cooperative voters cast 61\% of
their leadership votes for Subversive or Opposing candidates. Leaders then
removed the riders who followed the rules: the mean compliance rank of a rider
removed by a leader was 0.30 within its bike (0 is most compliant, 0.5 is
chance; Wilcoxon $p<.001$), while its score rank sat at chance (0.54;
Figure~\ref{fig:sanctions}b). Group votes operated differently: the rider voted
out was the bike's top scorer in 56\% of votes and its bottom scorer in 17\%
(mean score rank 0.31, $p<.001$), while effort (0.46) and compliance (0.46)
ranks remained at chance. Discussion matched: 57\% of turns referred to loot or score, 11\% to effort or pedalling. Score, however,
carried almost no information about contribution, because every rider aboard
is paid from every box (Spearman $\rho=-0.077$ with effort, $\rho=-0.05$ with
compliance). Collective sanctions targeted the one per-rider quantity the riders could see, and it did not reflect the behaviour the rules were about (measures defined in Appendix~S3.2).

\subsection{Hypotheses and limitations}

H1 is supported on membership and not on rules or regime. Every bike eliminated
riders, 302 times in all, but no bike amended its rules and regime choice was
stable at four group votes in five, so instability expressed itself through
membership rather than through collective choice. H2 is supported for the
physical measures and not for governance. A Cooperative plurality raised effort
and survival and a Lazy plurality lowered effort and nearly doubled Owdi
strikes, as predicted, but Subversive and Opposing pluralities did not produce
more leader regimes, more rule changes or more eliminations than the other
conditions. Composition changed how the bike moved, but did not change how it was governed.

Four limitations bound these claims. The design is between-session, so the error term is variation between sessions of one condition; effort is large against it, session length is not. Each condition doubles a disposition rather than
isolating it, so effects are those of increased prevalence in a mixed group.
Compliance and effort were set by each personality's movement controller, so
they are manipulations rather than findings; what the language model chose was
what to say, whom to nominate, whom to accuse and how to vote, and our
conclusions rest on those choices. Finally, the model was small: of 767 binding
turns naming a rider, 29 (3.8\%) carried a stated reason inconsistent with the
vote beside it.

\section{Discussion}
\label{sec:discussion}

Three findings came out of our initial demonstration. The first is constitutional inertia. Our
riders could amend their rules and never did. All 205 reopening votes failed,
even as loot fell five-fold between the first and the last round, where Scott
\emph{et al.} had found mutability to be the decisive factor in survival
\cite{scott2024megabike}. Deliberative energy went into membership, i.e., whom
to expel, rather than into collective choice; in Ober's vocabulary
\cite{ober2017}, entrenchment was treated as a single founding act. Second,
compliance and contribution came apart, and so did agreement and action. While Lazy
riders complied perfectly while pedalling at a seventh of maximum, Opposing
riders almost never complied while pedalling nearly as hard as Cooperative
ones. Moreover, Opposing bikes agreed to the most demanding effort rules while
realising the least effort. A metric that scores a group on its rate of
rule-following, as most evaluations of LLM collectives do
\cite{leibo2021meltingpot,pan2023machiavelli}, would observe none of this machine behaviour. Third,
sanctions missed their target in two opposite ways. Group votes removed the top
scorer, the only per-rider signal the riders could see and one unrelated to
contribution, which is a precise failure of Ostrom's monitoring principle
\cite{ostrom1990}. Leaders, elected most often from the riders least willing to be governed
and endorsed by Cooperative voters, removed the most compliant riders, which is
democratic backsliding on a short clock \cite{mertzani2024cohesion,pittober2018}.
In a hybrid society this is the type of attack on self-governance we have described before
\cite{sarkadi2024selfgov,piazza2025cpr}, where deliberation that is fluent but
indifferent to what it commits the speaker to \cite{hicks2024bullshit}, in
which the quantity of argument signalled neither its quality nor its sincerity.

Megabike differs from the main alternative for studying LLM collectives. Concordia \cite{vezhnevets2023concordia} is a library for generative agent-based modelling in which a Game Master, itself a language model, narrates the environment and adjudicates in natural language what the agents' free-text actions bring about. It is flexible and well suited to open-ended social simulation, but not fit for our kind of study in machine behaviour. In Megabike the environment, the institution and the sanctions are symbolic and executed. Energy, movement, compliance and expulsion are computed by the simulator, so whether a rule was followed does not depend on a second model's reading of a sentence, and cannot be argued with because it is simply grounded in a virtual environment. The agents also share one body, so a collective decision has consequences none of them can narrate away and defection costs the defector too. Moreover, a rider seat is agent-agnostic. The platform runs in a game engine in real time, and a human can take any seat in any session, which makes it a serious game for hybrid institutions rather than a text-based simulation of them. The price we pay with Megabike is a narrower space of actions. Riders can only do what the arenas in Table~\ref{tab:arenas} allow. Indeed, a meaningful testbed for self-governance has to fix what counts as a rule, a decision and a breach, or it cannot measure whether they are respected.

Using the testbed also showed us its current limitations, several of which are design choices we would now reverse. Transfers between bikes were disabled, so Ober's self-selection phase and the inter-bike bargaining it allows were switched off and every expulsion was an execution; with an exit, the balloon debate becomes a market and the leader's purge has a cost. The auditor role in Table~\ref{tab:arenas} was not instantiated, which is why riders had only score to go on, and the sanction ladder had one rung. Finally, 120-second rounds and five-round games give institutions little time to be tested by changing conditions, which is where mutability should pay off.

The immediate next steps are to make effort and compliance observable through an audit mechanism with a cost, to add graduated sanctions, to turn transfers on, and to repeat the experiment with larger models and with humans in some of the seats. We also propose a measure that this study makes available and that we think any claim of self-governance should report: the conditional probability that a sanction falls on a rider who broke the rule it enforces. In our experiment that probability was at chance under collective rule and below chance under a leader. A hybrid institution in which it is reliably above chance has not been demonstrated yet, and Megabike is a place to try.

\section{Conclusion}

In this paper, we introduced Megabike, a scenario and a testbed in which agents must share a single embodied vessel, agree on how to govern it, and survive under cooperative and adversarial
pressure, with all their rules and actions open to inspection.

The approach we described here provides researchers and
other users with tools to specify detailed models of both LLM-based and human behaviour in the context of socially constructed political rules that are inherently unpredictable given just the starting conditions. As with any research methodology it should be expected to come with its own strengths and limitations.


Megabike was built so that humans and machines can take the same
seats. As AI agents move from assisting people to deciding alongside them,
the institutions they form together will need to be tested before they are
trusted, and a testbed in which every rule is explicit and every action is
logged is where we can begin testing in a responsible, transparent, interpretable, and safe way.


\section*{Ethics}
This work did not involve human participants, personal data, or animal subjects, and required no ethical approval.

\section*{Data accessibility}
The forty session logs, the analysis code that produces every figure and table, and the electronic supplementary material are available at \url{https://zenodo.org/records/23661081}.

\section*{Author contributions}
C.B.: software, experiments, data analysis, writing; J.P.: conception, framework design, supervision, writing; S.S.: conception, experimental design, analysis, supervision, writing. All authors approved the final version.

\section*{Funding}
SS was supported by the Royal Academy of Engineering
and the Office of the Chief Science Adviser for National Security
under the UK Intelligence Community Postdoctoral Research Fellowship program.

\bibliographystyle{unsrtnat}
\bibliography{refs}

\clearpage
\appendix
\section*{Supplementary material}
\section*{S1\quad Agent configuration}

All ten riders in every session were controlled by Qwen3.5-4B-Q4\_K\_M running locally through the LLMUnity library inside the Unity game engine (Unity 6000.3.17f1, product ``MegaBike 2D''). 

\subsection*{S1.1 Personality prompts}

Each personality is set by one of the following system prompts.

\begin{itemize}
    \item \textbf{Cooperative}: You are cooperative and agreeable. You look for common ground, give people the benefit of the doubt, and tend to go with the group.
    \item \textbf{Lazy}: You are passive and low-effort. You rarely take initiative, your reasons are thin, and you often just follow whoever spoke last.
    \item \textbf{Subversive}: You are inconsistent and hard to read. Sometimes you cooperate, sometimes you undermine. Your stated reasons should sound plausible even when your position shifts.
    \item \textbf{Opposing}: You are contrarian and argumentative. You push back on what others want, challenge their reasoning, and resist going along with the group.
\end{itemize}

\subsection*{S1.2 Communication rules}

A common set of communication rules guided how agents addressed one another; these were produced through an iterative process to refine dialogue length and maintain focus on the game.

\begin{itemize}
    \item Keep responses to 1-2 sentences, under 35 words.
    \item Always say WHY you think what you think. Give a reason, not just a name.
    \item Base your reasons on what others have actually said in the discussion.
    \item Always speak in the FIRST PERSON as yourself.
    \item Never refer to yourself in the third person.
    \item Do not be overly enthusiastic.
    \item Do not use exclamation marks unless genuinely appropriate.
    \item Speak naturally, like a person arguing their case in a game.
    \item Do not repeatedly praise other riders.
    \item Do not talk about loot boxes, targets, collecting, or steering.
\end{itemize} 

\subsection*{S1.3 World view}

Each agent was given a small world view, enough to choose its next action:
\begin{itemize}
    \item its own identity and colour;
    \item every agent's loot score, with the leader marked;
    \item who has stated a leaning so far;
    \item the last twelve lines of dialogue;
    \item its permitted answer options.
\end{itemize}

\subsection*{S1.4 Answer options}

\section*{S2\quad Game parameters and workflow}

Each round lasts 120 seconds and a session lasts up to five rounds. A collected box gives one point to each rider aboard and one to the bike's pot, two points to a rider whose colour matches the box, and restores stamina to all riders. Each rider starts with 100 stamina. Steering costs no stamina; pedalling does. The Owdi crosses the arena on fixed linear paths and forces the bike it touches to remove a rider; a rider left alone on a bike that the Owdi catches is eliminated. Transfers between bikes were disabled.

\section*{S3\quad Statistical methods}

Because experimental conditions changed from session to session, we compared the four conditions using a Kruskal--Wallis test across the ten sessions per condition. However, all four personality types were present in every session, making personality comparisons within-session. We used Holm's method to adjust for multiple comparisons. When we picked a specific comparison to test after seeing the overall results, we checked it against a randomized permutation distribution. Any result with $p < .001$ remains statistically significant regardless of the correction method used.

Condition-level tests are Kruskal--Wallis on four groups of ten sessions ($df=3$ throughout). Personality-level measures are within-session and computed on session-level means; pairwise comparisons use paired Wilcoxon tests with Holm correction. Contrasts chosen after inspection are checked against a randomised permutation distribution and reported at their selection-adjusted significance.

\subsection*{S3.1 Power for the elimination null}
Each personality occupies exactly 100 rider-slots, so elimination counts compare directly: Cooperative 76, Lazy 83, Subversive 69, Opposing 74. A $2\times4$ test does not reject equality ($\chi^2=5.46$, $df=3$, $p=.141$), a session-stratified permutation test agrees ($p=.167$), and the round of elimination was undifferentiated ($H=2.76$, $p=.430$); the widest pair, Lazy against Subversive, is nominally significant alone (Fisher $p=.031$) but not after correction ($p=.184$). This null is not equivalence: at 100 slots and a base rate near 75\%, the design had 80\% power only against a shift to roughly 90\%, and the Wilson intervals overlap heavily. The defensible claim is therefore narrow: Opposing and Cooperative riders were eliminated at 74\% and 76\% despite differing in compliance by 98.6 percentage points.

\subsection*{S3.2 Elimination hazard and within-bike rank}
The elimination hazard in Figure~\ref{fig:sanctions}a is defined per deliberation: the denominator for a personality is the number of its appearances in a deliberation that reached the elimination stage (a group vote or a leader's decision), and the numerator is the number of those appearances that ended in its removal. Pleas to a leader and binding votes identify which riders were present. Intervals are 95\% Wilson intervals. The within-bike rank in Figure~\ref{fig:sanctions}b is the removed rider's position among the riders aboard when all are sorted on the signal in the preceding round, scaled to $[0,1]$ with 0 the highest; 0.5 is chance, and the test against chance is a one-sample Wilcoxon on the centred ranks. The sanctions analysis is exploratory and its tests are not corrected for multiplicity.

\section*{S4\quad Discourse analysis}

\subsection*{S4.1 How riders argue: full account}
 
Through analysing the 415 deliberations, three features were observed. First, the deliberations are not about the discussion trigger event. Only 1.7\% of Owdi removal discussions refer to the collision, the vehicle or any threat. This is marginally less than where no collision occurred at 2.6\%. Second, a rider's collected loot is argued as both a positive and a negative. Of 1,632 turns mentioning a rider's loot standing, 250 (15\%) treat it as a negative for the team and 166 (10\%) treat it as a positive, with 60 using both at once. Because a score reflects what the whole bike collected while a rider was aboard, it supports a narrative of greed and contribution. Agents also accuse each other of impossible acts, such as lying (109 turns), theft (80), framing or conspiracy (23), and sabotage (13), and they read outputs of the allocation rule as evidence of collusion: \textit{``their perfect score match with Player 9 makes them suspicious of trying to frame me''} (Session 16, Deliberation 3).
 
Third, agents will attach to targets already chosen rather than to new agent behaviour. Silence is the clearest case of this where 192 turns cite a named agent's silence, yet of the 125 that name an agent during discussion, 73 name an agent who had already spoken in that same discussion turn. Accusation is also highly reciprocal. Of 3,029 discussion turns, 32\% name an agent who has already named the speaker, where 23.4\% would be expected from the targeting rates alone. Additionally, under leader governance, the agent removal decision of the leader frequently differs from their original choice in 38\% of deliberations (38 of 101).

Rule voting outcomes were approved with almost no exception. Of 611 individual ballots on whether to reopen the agreed rules, 607 were against, so none of the 205 collective reopening votes passed and every bike played to the end under the rules it adopted in its first deliberation. The agents' justifications treat whatever happened in the previous round as validation that the rules work, including an agent removal the speaker knows was settled by a coin flip (Session 29, Deliberation 5).

\subsection*{S4.2 Lexical registers}
Looking towards the common words used by each personality, cooperative uses \textit{agree} and \textit{fairer}, Subversive uses \textit{shifting} and \textit{switching}, Opposing uses \textit{lie}, \textit{greedy}, \textit{blindly}, and Lazy uses \textit{since}, \textit{said}, \textit{argued}. The high justification rate in Lazy agents (83\%) is mostly citation of other agents than its own argument. As an example, a rider saying \textit{``I'll go with Group since Player 8 just said trusting one person is risky''} is considered as agreeing and reasoning while providing little contribution. Scoring each word by its log-odds ratio against the other three personalities gives each a distinct register: Cooperative \textit{agree} (492 uses) and \textit{fairer}; Subversive \textit{shifting} and \textit{switching}; Opposing direct insult, \textit{lie}, \textit{greedy} and \textit{blindly} (56 uses against six or fewer elsewhere); Lazy \textit{since} (800 uses), \textit{said} and \textit{argued}.

\section*{S5\quad Stated versus enacted position}

The second is a weak correspondence between an agent's initial stated position and its final binding vote. A rider's opening discussion leaning matched their binding vote in 41.4\% of cases and their final leaning in 43.1\%. Because votes typically converge on a single target, 27.1\% of this agreement arises by coincidence alone, so discussion does predict the vote above chance ($p<.001$). However, a majority of that prediction is provided in an agent's first leaning, while the deliberation that follows it adds under 2\% difference.

Differences between personalities here are largely due to abstaining to vote. Discussion offers no ``Nobody'' option, so every leaning names a rider, while 14.8\% of binding votes were ``Nobody'' and therefore could not match. Abstention was personality-dependent (Cooperative 3.1\%, Subversive 12.1\%, Lazy 21.0\%, Opposing 25.3\%). Restricted to votes naming a rider, correspondence is 57.2\% for Cooperative, 53.4\% for Lazy, 53.2\% for Subversive and 36.0\% Opposing. The Opposing personality argued most forcefully and were the only group whose stated positions failed to predict their votes.

\section*{S6\quad Data and code}

The forty session logs (agents, events, utterances, movement and collection tables per session) and the analysis scripts that produce every figure and table are deposited at: https://doi.org/10.5281/zenodo.23661081

\end{document}